\documentstyle[11pt]{article}
\makeatletter

\gdef\@publabel{\hfil}
\gdef\@pubdate{\null}
\gdef\@pubnumber{\null}
\gdef\@author{\null}
\gdef\@title{\null}
\gdef\@abstract{\null}
\long\def\pubdate#1{\gdef\@pubdate{#1}}
\long\def\pubnumber#1{\gdef\@pubnumber{#1}}
\long\def\publabel#1{\gdef\@publabel{#1}}
\long\def\author#1{\gdef\@author{#1}}
\long\def\title#1{\gdef\@title{#1}}
\long\def\abstract#1{\gdef\@abstract{#1}}
\def\titlerelax{
\let\maketitle\relax
\let\settitleparameters\relax
\let\consolidatetitle\relax
\let\inittitlepage\relax
\let\finishtitlepage\relax
\let\titlepagecontents\relax
\let\multithanks\relax
\let\titlebaselines\relax
\let\@makepub\relax
\let\@maketitle\relax
\let\@makeauthor\relax
\let\@makeabstract\relax
\let\@maketitlenote\relax
\let\thanks\relax
\let\titlerelax\relax}
\def\titleclean
{\gdef\@titlenote{}
\gdef\@abstract{}
\gdef\@author{}
\gdef\@title{}
\gdef\@pubdate{}\gdef\@pubnumber{}\gdef\@publabel{}
\gdef\@dpublabel{}
}

\def\@makepub{\vbox to \z@{\hbox to \textwidth{\hfill
\@publabel \hfill
\llap{\parbox[t]{0.33\textwidth}{\raggedleft\@pubnumber}}}%
\vss}}
\def\@maketitle{\vskip 60pt \begin{center}
 {\LARGE \@title \par}
 \end{center}}

\def\@makeauthor{{%
\def\and{\smallskip {\normalsize \rm and\smallskip }}
\def\And{\medskip {\normalsize \rm and\\}\medskip}
\long\def\address##1{{\def\and{\\and\\}\medskip
				{\small  \\##1\\}
}}
{\centering
 \vskip 3em
 \large \lineskip .75em
 \@author}
 \par}}
\def\@makedate{\vskip 1.5em
 {\raggedright \small \noindent\@pubdate \par}}
\def\@makeabstract{\vskip 1.5em
{\small
\begin{center}
{\bf ABSTRACT\vspace{-.5em}\vspace{0pt}}
\end{center}
\quotation \@abstract \endquotation}}
\def\maketitle{\titlepage
\let\footnotesize\small \setcounter{page}{0}
\@makepub
\vfil
\@maketitle
\@makeauthor
\vfil
\@makeabstract
\@thanks
\vfil
\@makedate
\if@restonecol\twocolumn \else \eject \fi
\titlerelax \titleclean
\setcounter{footnote}{0}
}

\pubnumber{UCLA/93/TEP/28}
\pubdate{July 1993}
\title{Intermittency in Branching Processes}

\author{Antonio O.\ Bouzas$^{1}$
\address{Department of Physics,\\ University of California, \\
Los Angeles, California 90024-1547}
}
\footnotetext[1]{Internet: bouzas@physics.ucla.edu}

\abstract{We study the intermittency properties of two branching processes, one
with a uniform and another with a singular splitting kernel. The
asymptotic intermittency indices, as well as the leading corrections
to the asymptotic linear regime are explicitly computed in an analytic
framework. Both models are found to possess a monofractal spectrum
with  $\varphi_{q}=q-1$. Relations with previous results are discussed.
}

\newcommand{\bea}{\begin{eqnarray}}
\newcommand{\eea}{\end{eqnarray}}
\newcommand{\be}{\begin{equation}}
\newcommand{\ee}{\end{equation}}
\newcommand{\wt}[1]{\widetilde{#1}}

\begin{document}

\maketitle
\section{Introduction.}

The original proposal of intermittency in multiparticle production
\cite{bialpes} consisted of a method of quantifying and measuring
short-range correlations in multi-hadron final states, as well as a
hypothesis about the self-similar nature of fluctuations. This fractal
character is supposed to have dynamical origins and expected to provide
valuable information about the underlying mechanism of hadron
production. The subject of intermittency has been intensely pursued
since its introduction in \cite{bialpes}, both theoretically
\cite{bialrev,pesrev} and experimentally
\cite{schmitz,buschbeck} and is by now a standard topic in soft
hadronic physics.

On the theoretical side, the property of intermittency was found to
hold in many models, most remarkably $\alpha$-models \cite{bialpes},
of which detailed studies have been made (\cite{brax,zal} and
references therein). A different class of systems where intermittency
is expected to be present is branching processes. These have been
applied to the phenomenological description of high-energy processes
both of electromagnetic \cite{bhar} and hadronic
\cite{polya,orfa,giova} nature.
In the context of intermittency, branching processes provide a
laboratory to study short-range correlations and fluctuations in
multi-particle production either analytically or numerically.

Intermittency in branching processes has been studied in \cite{hwa},
where a proof of intermittent behavior is offered. It has also been
considered in the more specific framework of QCD branching in
\cite{brax1,brax2}, where the problem of infrared divergences arises.
In this paper we adopt a point of view complementary, in some sense,
to that of \cite{brax1,brax2} and closer to \cite{hwa}. We shall study
the intermittent regime of two one-species branching models,
characterized by a uniform and a singular splitting kernel,
respectively. Our aim is to determine whether these mathematical
models display intermittent behavior, of what kind, in what limit, and
with what type of corrections. We are also interested in the
relationship \cite{pesc} between intermittency and KNO scaling \cite{koba}.
Our mathematical approach is different and, we believe, simpler than
that of \cite{hwa}, which allows us to obtain more specific results.
The price to pay for those results is that we have to give up
generality, by considering two particular cases,
and that we defer for future consideration the problem of
infrared-divergent kernels.

The outline of the paper is as follows. In Section 2, we define the
model, explain those general features of branching processes which are
relevant to our purposes, and fix our notations and conventions. In
Section 3, we study the particular models in terms of evolution
equations for inclusive distributions. For both splitting kernels, a
monofractal spectrum with maximal intermittency indices \cite{bialzal}
is found, as well as the leading corrections to the asymptotic
behavior of scaled factorial moments and a precise characterization of
the regime where intermittency appears. In section 4 , a summary of
results is given, together with our final remarks. An
Appendix, finally, gathers some complementary material related to the
mathematical approach used in the main body of the paper.

\section{Model. Notations and Conventions.}

The branching process under consideration consists of particles
characterized by an energy fraction $x$ and a virtuality $t$, the
latter being the evolution parameter of the system. Mathematically it
is defined by the splitting kernel $P(x)$ entering the evolution
equations for the probability $\cal P$ to having a branching at
virtuality $t$,
\be
\frac{d{\cal P}}{dt} = \int dx P(x)
\ee
This equation is inspired in the Altarelli-Parisi-Gribov-Lipatov
equations for partonic evolution \cite{alta,gribo}, with the
simplificatory assumption that the running coupling constant
$\alpha_{s}(Q^2)$ does not depend on $x$, $Q^2=Q^2(t)$, so that it can
be absorbed in the evolution variable $t$ \cite{webber,cvitanovic}. In
this way, a stationary Markov branching process is defined and
evolution equations for the generating functional of transition
probabilities can be found.  We shall omit the derivation here, which
is given in detail in, {\em e.\frenchspacing
g.\frenchspacing,} \cite{webber,cvitanovic,sjostrand} (see also the
Appendix, where the relevant equations are summarized).

We shall be interested, as will be made clear later in this section,
in inclusive transition probabilities $D_{n}(x_{1},\ldots,x_{n};t)$
which represent the probability to observing particles with energy
fractions $x_{1},\ldots,x_{n}$, as well as particles with any other
values of $x$, at virtuality $t$, and starting with one particle with
$x=1$ at $t=0$. These inclusive distributions are normalized to the
factorial moments of multiplicity,
\be\label{eq:2}
\int_{0}^{1} dx_{1}\ldots dx_{k} D_{k}(x_{1},\ldots,x_{k};t) =
                 \langle n (n-1)\cdots (n-k+1)\rangle (t)
\ee
where the multiplicity $n$ is the total number of particles in the system
at virtuality $t$. This normalization condition follows naturally from the
definition of $D_{n}$ in terms of the generating functional of the process
(see Appendix and \cite{webber,cvitanovic}).

We shall further assume that the spliting kernel $P(x)$ is normalized
to unity, which implies no loss of generality in the one-species case;
that it is symmetric ($P(x) = P(1-x)$), which follows from energy
conservation; and that its support is contained in the interval
$[0,1]$. Moreover, we shall adopt the convention that all probability
densities, inclusive or exclusive of any order, vanish outside
$[0,1]$.

The inclusive densities of order $n$, $D_{n}(x_{1},\ldots,x_{n},t)$,
satisfy the forward equation (see Appendix),
\bea
\lefteqn{\frac{\partial D_{n}}{\partial t}(x_{1},\ldots,x_{n},t) =}
\nonumber \\
& &\hspace{-4ex}     - n D_{n}(x_{1},\ldots,x_{n},t)
+ 2 \int_{0}^{1}\!\!\! dz\: P(z) \sum_{k=1}^{n} \frac{1}{z}
D_{n}(x_{1},\ldots,\frac{x_{k}}{z},\ldots,x_{n},t) \nonumber \\
& &\hspace{-4ex}\mbox{} + 2\sum_{k>j=1}^{n}
P\left(\frac{x_{j}}{x_{j}+x_{k}}\right)
\frac{1}{x_{j}+x_{k}} D_{n-1}(x_{1},\ldots,\underbrace{x_{j}+x_{k}}_{j},
\ldots,\widehat{x_{k}},\ldots,x_{n},t)
\label{eq:for}
\eea
where the hat indicates that the variable $x_{k}$ is omitted and we
used the symmetry of $P(z)$.
The initial condition is given by,
\be
D_{n}(x_{1},\ldots,x_{n},t=0)=\delta_{1n}\delta(x_{1}-1)
\ee

We now define the factorial moment densities of order $n$,
(cf.\frenchspacing,  eq.\
(\ref{eq:2}))
\be
d_{n}(x,t) = \int_{0}^{1}\!\!\! dx_{2}\cdots dx_{n}\:
D_{n}(x,x_{2},\ldots,x_{n},t)
\ee
Separating $x_{1}$ in the equation for $D_{n}$ and integrating
 $dx_{2}\cdots dx_{n}$ we obtain the forward evolution equation
for $d_{n}$,
\bea
\frac{\partial d_{n}}{\partial t}(x,t) & = & (n-2)
d_{n}(x,t)+(n-1)(n-2)d_{n-1}%
(x,t) \nonumber \\
& & \hspace{-2ex}\mbox{} + 2\int_{0}^{1}\!\!\! dz\: P(z)\frac{1}{z}
                                              d_{n}\left(\frac{x}{z}\right)
+2(n-1)\int_{0}^{1}\!\!\! dz\: P(z)\frac{1}{z} d_{n-1}\left(%
                                          \frac{x}{z}\right)
\label{eq:4}
\eea
with  initial condition,
\be
d_{n}(x,t=0)  =  \delta_{1n}\delta(x-1)
\ee
The Mellin transformed equation,
\bea
\frac{\partial\widetilde{d}_{n}}{\partial t}(s,t) & = &
\left( n-2+2\widetilde{P}(s)\right)\left(\widetilde{d}_{n}(s,t)+
(n-1)\widetilde{d}_{n-1}(s,t)\right) \\
\widetilde{d}_{n}(s,t=0) & = & \delta_{1n}
\eea
can be explicitly solved, to obtain,
\be
\widetilde{d}_{n}(s,t) = e^{ (n-2)t} \left( 1 -
e^{ -t}\right)^{ n-1} e^{  2\widetilde{P}(s)%
   t}  \:\frac{\Gamma\left(n-1+2 \widetilde{P}(s)\right)}{\Gamma\left(2
\widetilde{P}(s)\right)}
\ee
Substituting $s=1$ in this expression, we obtain the factorial moments
corresponding to a pure-birth, binary-fission process, which are
independent of the
form of the splitting kernel \cite{bouzas},
\be
\langle n(n-1)\cdots (n-q+1)\rangle = q! e^{   qt} (1-e^{  -t})^{  q-1}
\ee
We shall find useful later the following expanded form of the solution,
\be \label{eq:10}
\widetilde{d}_{n}(s,t) = e^{ (n-2)t}\left( 1 -
e^{ -t}\right)^{ n-1} e^{  2\widetilde{P}(s)%
   t}\sum_{m=1}^{n-1} C_{m}^{n-1}\left( 2\widetilde{P}(s)\right)^{m}
\ee
where
\be
\left\{
\begin{array}{rcl}
C_{n-1}^{n-1} & = & 1\\
C_{m}^{n-1} & = &  \sum_{i_{1}>\cdots >i_{r}=1}^{n-2} i_{1}\cdots i_{r}
\label{eq:cm}
\end{array}
\right.
\ee
with $r  =  n-1-m $ and $n>2$.

\subsection{Intermittency indices.}
\label{sec:goal}

Our goal in this paper is to obtain the intermittency indices of a
branching process, defined as the logarithmic slope of reduced
factorial moments as functions of bin size \cite{bialpes}. Reduced
factorial moments are defined as,

\be
F_{q} \equiv \frac{U_{q}}{U_{1}^{q}}
\ee
$U_{q}\equiv \langle n\cdots (n-q+1)\rangle$ being the factorial
moment of the multiplicity within the given bin.  Notice that
factorial moments $U_{q}$ are obtained from the above-defined
densities as,
\be
U_{q}(x;t) = \int dx\:  d_{q}(x;t)
\ee
where the integration extends to the bin under consideration.

Due to the asymmetry of the process, $\langle x\rangle (t)$ being a
monotone decreasing function of $t$ (see Appendix), the relevant bin in any
partition
of $[0,1]$ is the one adjacent to the point $x=0$. For that reason, it
is appropriate to apply vertical analysis (see, {\em e.\frenchspacing
g.\frenchspacing,} \cite{schmitz,pespes}) to that bin in order to find the
intermittency slopes.  However, the particle density $U_{1}$ need
not be uniform within the bin under consideration, so that we shall
define  intermittency indices in terms of $F_{q}$ as a function of
$U_{1}$ \cite{ochs},
\be
\frac{d\ln F_{q}}{d\ln U_{1}} = \frac{\frac{d\ln U_{q}}{dx}}%
{\frac{d\ln U_{1}}{dx}} - q \equiv -\varphi_{q}
\ee
This is the definition of intermittency indices that we shall adopt in
the sequel.

Before going to a particular case in the next section, we would like
to mention another consequence of the asymetry of the process.  For
long enough $t$, and a fixed resolution in $x$, almost all particles
will be contained in the bin adjacent to $x=0$, where the process will
continue its evolution as a usual branching process with only
multiplicity degrees of freedom. But this kind of processes KNO scale
\cite{orfa,koba,bouzas},
leading to constant reduced factorial moments in the limit
$t\rightarrow \infty$, and consequently to zero intermittency
indices in that limit. We shall explicity verify this fact below .

\section{Two Particular Cases.}

In this section, we shall compute  the asymptotic form of
reduced factorial moments with the expressions derived in the previous
section. Factorial moment densities can be exactly found for
$P(x)=1$ and $P(x)=\delta(x-1/2)$, which we shall exploit to obtain
the asymptotic moments at a given resolution.

\subsection{Uniform Splitting Kernel.}

For the ``uniform model'', we have $P(z)=1$,
$\wt{P}(s)=1/s$.  By expanding the exponential in eq. (\ref{eq:10}),
we find
\bea
\wt{d}_{1}(s,t) & = & e^{  -t}\sum_{n=0}^{\infty}\frac{(2t)^{  n}}{n!}
		\frac{1}{s^{ n}}\\
\wt{d}_{2}(s,t) & = & 2(1-e^{  -t})\sum_{n=0}^{\infty}\frac{(2t)^{  n}}{n!}
		\frac{1}{s^{ n+1}}\\
\wt{d}_{n}(s,t) & = & e^{  (n-2)t}(1-e^{  -t})^{  n-1}\sum_{m=1}^{n-1}
		C_{m}^{n-1}2^{ m}\sum_{k=0}^{\infty}\frac{(2t)^{
k}}{k!} 		\frac{1}{s^{ k+m}}
\eea
Applying the inverse Mellin transform term by term leads to,
\bea
d_{1}(x,t) & = & e^{ -t}\sqrt{\frac{2t}{-\ln x}}\:
I_{1}\left(\sqrt{-8t\ln x}\right) + e^{ -t}\delta(x-1)\\ d_{2}(x,t) &
= & 2(1-e^{ -t}) I_{0}\left(\sqrt{-8t\ln x}\right) \\ d_{n}(x,t) & = &
2e^{ (n-2)t}(1- e^{ -t})^{ n-1}\times\nonumber\\ & \times
&\sum_{m=1}^{n-1} C_{m}^{n-1}%
\left(\sqrt{\frac{-2\ln x}{t}}\right)^{  m-1} I_{m-1}\left(\sqrt{-8t\ln
x}\right)
\eea
where $I_{n}$ denotes a modified Bessel function of the first kind and
order $n$ \cite{abra}.  Since these expressions are given by finite
sums, in order
to check that they are solutions to eq.\ (\ref{eq:4}) it is enough to
see that their Mellin transform is correct. This is easily done by
change of variables, turning one-sided-Mellin into Laplace transforms
and using the known transforms of Bessel functions \cite{erdelyi}.

The factorial moments $U_{n}(x,t)\equiv\int_{0}^{x} dz\; d_{n}(z,t)$
are then given by
\bea
U_{n}(x,t) & = & 2e^{ (n-2)t}(1- e^{ -t})^{ n-1}\times\nonumber\\ &
\times &\sum_{m=1}^{n-1} C_{m}^{n-1}\int_{-\ln x}^{\infty}\!\!\! dy\:
e^{ -y}\left(\sqrt{\frac{2y}{t}}\right)^{ m-1}
I_{m-1}\left(\sqrt{8ty}\right)
\eea
and analogously for $U_{1}$, $U_{2}$.  We shall now assume that the
argument of the Bessel functions is large, so that we can approximate
them by their asymptotic expansion \cite{abra} to obtain,
\bea
U_{n}(x,t) & = & 2e^{ (n-2)t}(1- e^{ -t})^{ n-1}\sum_{m=1}^{n-1}
C_{m}^{n-1} 	 \int_{-\ln x}^{\infty}\!\!\! dy\:e^{%
-y}\times\nonumber\\
& \times &\left(\sqrt{\frac{2y}{t}}\right)^{ m-1}
\frac{e^{ \sqrt{8ty}}}{\sqrt{2\pi\sqrt{8ty}}}\left(1-\frac{4(m-1)^{
2}-1}{8\sqrt{8ty}} + \cdots\right) \label{eq:20}
\eea
where we have retained the next-to-leading term in order to estimate
the order of magnitude of corrections later.

\subsubsection{Large $t$.}

We consider now a resolution $x$ as small as desired but fixed, and
$t\rightarrow\infty$.  We can then drop the correction term in eq.\
(\ref{eq:20}) and re-write the leading term by means of the
substitution $\eta=(y/2t)^{1/4}$ as,
\bea
U_{n}(x,t) & = & 2e^{ nt}(1- e^{ -t})^{ n-1}\sum_{m=1}^{n-1}
C_{m}^{n-1} 2^{ m} \sqrt{\frac{2t}{\pi}}\times\nonumber\\ & &\times
\int_{\left(\frac{-\ln x}{2t}\right)^{1/4}}^{\infty}\!  d\eta\: e^{
-2t(\eta^{2}-1)^{2}} \eta^{ 2m}
\eea
For $t\rightarrow\infty$ we can estimate this expression by saddle
point in the form,
\bea
U_{n}(x,t) & \approx & 2e^{ nt}(1- e^{ -t})^{ n-1}\sum_{m=1}^{n-1}
C_{m}^{n-1} 2^{ m} \sqrt{\frac{2t}{\pi}}\times\nonumber\\ & &
\times\int_{\left(\frac{-\ln x}{2t}\right)^{1/4}}^{\infty}\!\!\!
d\eta\: e^{ -8t(\eta-1)^{2}} \eta^{ 2m}
\eea
or, for $\rho = \sqrt{8t}(\eta-1)$,
\bea
U_{n}(x,t) &\approx & 2e^{ nt}(1- e^{ -t})^{ n-1}\sum_{m=1}^{n-1}
C_{m}^{n-1} 2^{ m-1} \sqrt{\frac{1}{\pi}}\times \nonumber\\ & &
\times\int_{-\infty}^{\infty}\!\!\!  d\rho\: e^{ -\rho^{2}}
\left(\frac{\rho}{2\sqrt{2t}}+1\right)^{ 2m}
\eea
Notice that we substituted $-\infty$ for $-\sqrt{8t}$ in the lower
limit of the integral.  Applying the binomial expansion to the bracket
in the integrand and integrating term by term, the resulting sum shows
that in the limit the integral takes the value $\sqrt{\pi}$. Then, by
definition of $C_{m}^{n-1}$, eq.\ (\ref{eq:cm}), we find $U_{n}(x,t)
\approx n! e^{ nt}$, which corresponds to the KNO value given by,
\be
\frac{U_{n}}{U_{1}^{n}} = n!   \;\;\; (t\rightarrow\infty)
\ee
As discussed in the previous section, this leads to zero intermittency
indices.

\subsubsection{Small $x$.}

For $t$ fixed, we now consider the regime of small $x$. Although the
scale of smallness will be apparent as we proceed with the computation
of factorial moments, we can in principle assume $x\ll \langle
x\rangle$, where $\langle x\rangle = t/(e^{t}-1)$ is the average
energy fraction (see Appendix).

By changing the integration variable in eq.\ (\ref{eq:20}) to
$\eta=\sqrt{y}-\sqrt{2t}$, we obtain,
\bea
U_{n}(x,t) & = & e^{ nt}(1- e^{ -t})^{
n-1}\frac{4}{\sqrt{2\pi}(8t)^{1/4}} \sum_{m=1}^{n-1}
C_{m}^{n-1}\left(\frac{2}{t}\right)^{ \frac{m-1}{2}}
\times\nonumber\\ & & \times\int_{a}^{\infty}\!\!\! d\eta\:e^{
-\eta^{2}} 	 \eta^{ m-1/2} \left(
1+\frac{\sqrt{2t}}{\eta}\right)^{ m-1/2} \times\nonumber\\ 	 &
&\times\left( 1-\frac{(m-1)^{2}-1}{4\sqrt{2t}\eta\left(
1+\frac{\sqrt{2t}}{\eta} 	 \right)}+\cdots\right)
\eea
where we denoted $a\equiv\sqrt{-\ln x}-\sqrt{2t}$ for brevity.  By
applying the series expansion, valid for $a>\sqrt{2t}$,
\bea
\lefteqn{\int_{a}^{\infty}\!\!\! d\xi\: \xi^{
p/2}\left(1+\frac{\sqrt{2t}}{\xi}\right)^{  p/2}
e^{ -\xi^2} = } \hspace{5ex} \nonumber\\ & =
&\sum_{n=0}^{\infty}\frac{1}{2n!}\frac{\Gamma(p/2+1)}{\Gamma(p/2-n+1)}
\left(\sqrt{2t}\right)^{
n}\Gamma\left(\frac{p}{4}-\frac{n}{2}+\frac{1}{2},a^2\right)
\eea
to the previous expression of $U_{n}$, we have,
\bea
U_{n}(x,t) & = & e^{ nt}(1- e^{ -t})^{
n-1}\frac{4}{\sqrt{2\pi}(8t)^{1/4}} \sum_{m=1}^{n-1}
C_{m}^{n-1}\left(\frac{2}{t}\right)^{ \frac{m-1}{2}}
\times\nonumber\\ 	 & \times &
\sum_{k=0}^{\infty}\frac{\left(\sqrt{2t}\right)^{ k}}{2k!}
\frac{\Gamma(m+1/2)}{\Gamma(m+1/2-k)}
\Gamma\left(\frac{m}{2}-\frac{k}{2}+\frac{1}{4},a^2\right)
\times\nonumber\\ 	 &\times &\left[1-\frac{(m-1)^{
2}-1}{8t}\frac{k}{m-1/2}+\cdots\right]
\eea
Some remarks about this expression are in order here,
\begin{trivlist}
\item[i.] The correction factor in brackets comes from the asymptotic expansion
of
the modified Bessel function. Since it is linear in $k$, it does not
affect the rate of convergence of the series as given, {\em
e.\frenchspacing g.\frenchspacing ,} by the ratio test.
\item[ii.] This factor does not affect the zeroth-order term $k=0$. Similarly,
the next
order term would not affect the $k=0$ and $k=1$ terms, {\em etc.}
\item[iii.] As we shall see, only the leading term in the series need be kept.
If we
wanted to retain more terms in the series, we would have to take into
account the next corrective terms coming from the asymptotic expansion
of the Bessel function.  Such expansion is, however, divergent, so
that only a finite number of terms may be kept.
\end{trivlist}
Regarding the convergence of the series, we can use the following
asymptotic expression, obtained by a saddle-point argument for large
$\omega$,
\be
\Gamma(-\omega,x) = \frac{e^{  -x} x^{  -\omega}}{\omega+x+1} [1+\cdots]
\;\;\;\;\;\;\;\;\;(\omega\rightarrow +\infty)
\ee
to obtain the rate of convergence of the series as
\be
\left|\frac{u_{n+1}}{u_{n}}\right| \rightarrow \frac{\sqrt{2t}}{\sqrt{-\ln
x}-\sqrt{2t}}
\ee
We see, then, that the series is absolutely convergent for $8t<-\ln x$
and, furthermore, that for $\sqrt{2t}\ll \sqrt{-\ln x}$ it is
justified to truncate it.  We shall retain the leading ($k=0$) and
next-to-leading ($k=1$) terms, the second one in order to have an
estimate of the corrections.
\bea
\lefteqn{U_{n}(x,t)  \cong }\nonumber\\
& &e^{ nt}\left(1-e^{ -t}\right)^{
n-1}\frac{4}{\sqrt{2\pi}(8t)^{1/4}}\times\nonumber\\ &
&\times\sum_{m=1}^{n-1} C_{m}^{n-1}\left(\frac{2}{t}\right)^{
\frac{m-1}{2}}\frac{1}{2}
\Gamma\left(\frac{m}{2}+\frac{1}{4},a^2\right)\times\nonumber\\
& &\times\left[
1+\sqrt{2t}(m-1/2)\frac{\Gamma\left(\frac{m}{2}-\frac{1}{4},a^2\right)}{
\Gamma\left(\frac{m}{2}+\frac{1}{4},a^2\right)}
\left( 1-\frac{(m-1)^{2}-1}{8t}\frac{1}{m-1/2}\right)\right]
\eea
We now use the asymptotic form \cite{abra} for $\omega > 0$ fixed,
$x\rightarrow\infty$,
\be
\Gamma(\omega,x) = e^{  -x} x^{  \omega -1}\left( 1+\frac{\omega -1}{x}
+\cdots\right)
\ee
to obtain
\bea
U_{n}(x,t) & \cong &e^{ nt}\left(1-e^{ -t}\right)^{
n-1}\frac{4}{\sqrt{2\pi}(8t)^{1/4}}\times\nonumber\\ &
&\times\sum_{m=1}^{n-1} C_{m}^{n-1}\left(\frac{2}{t}\right)^{
\frac{m-1}{2}}\frac{1}{2}
\left(e^{  -a^2} a^{  m-3/2}\right)\times\nonumber\\
& &\times\left[ 1+(m-1/2)\frac{\sqrt{2t}}{a}
\left( 1-\frac{(m-1)^{2}-1}{8t}\frac{1}{m-1/2}\right)\right]
\eea
or, recalling that $a\equiv\sqrt{-\ln x} - \sqrt{2t}$,
\bea
U_{n}(x,t) &\cong & e^{ (n-2)t}\left(1-e^{ -t}\right)^{ n-1}\frac{2e^{
\sqrt{-8t\ln x}}x}{\sqrt{2\pi}(8t)^{1/4}}\times\nonumber\\ &
&\times\sum_{m=1}^{n-1}
C_{m}^{n-1}\left(\frac{2}{t}\right)^{\frac{m-1}{2}} (-\ln
x)^{\frac{m}{2}-\frac{3}{4}}\times\nonumber\\ & &\times\left[
1+\sqrt{\frac{2t}{-\ln x}}
\left( 1-\frac{(m-1)^{2}-1}{8t}\right)\right]
\eea
The leading contribution in the limit $x\rightarrow 0$ comes from the
$m=n-1$ term. Keeping only this and the next-to-leading one we finally
have,
\bea
\lefteqn{U_{n}(x,t)  =} \nonumber\\
& & \left( e^{(n-2)t}\frac{\left(1-e^{ -t}\right)^{n-1}}{(8t)^{1/4}}
\left(\frac{2}{t}\right)^{ \frac{n}{2}-1}\right) \sqrt{\frac{2}{\pi}}
\left(e^{\sqrt{-8t\ln x}}x(-\ln
x)^{\frac{n}{2}-\frac{5}{4}}\right)\times\nonumber\\
& &\times \left[ 1+\sqrt{\frac{2t}{-\ln
x}}\left(1-\frac{(n-2)^{2}-1}{8t} +
\frac{(n-1)(n-2)}{4}\left(\frac{2}{t}\right) ^{
\frac{n}{2}-1}\right)\right]
\eea

\paragraph{Mean value $U_{1}$ and factorial moment $U_{2}$.}

For the cases $n=1$, 2 we have,
\bea
U_{1}(x,t) & = & e^{ -t}\int_{0}^{x}\!\!\! dx\: I_{1}(\sqrt{-8t\ln
x})\sqrt{\frac{2t}{-\ln x}} \\ U_{2}(x,t) & = & 2(1-e^{
-t})\int_{0}^{x}\!\!\! dx\: I_{0}(\sqrt{-8t\ln x})
\eea
Expanding the Bessel functions as before, we arrive at
\bea
U_{1}(x,t) & = & \frac{e^{t}
(8t)^{1/4}}{\sqrt{2}}\sum_{n=0}^{\infty}\frac{1}{2n!}\frac{(\sqrt{2t})^{n}}{\Gamma(1/2-n)}
\Gamma\left(\frac{1}{4}-\frac{n}{2},a^2\right)\left[1+\frac{3n}{2t}\right] \\
U_{2}(x,t) & = &
\frac{(1-e^{t})e^{2t}}{(2t)^{1/4}}\sum_{n=0}^{\infty}\frac{1}{2n!}\frac{(\sqrt{2t})^{n}}{\Gamma(3/2-n)}
\Gamma\left(\frac{3}{4}-\frac{n}{2},a^2\right)\left[1+\frac{n}{16t}\right]
\eea
Retaining only the leading and next-to-leading terms and expanding the
gamma functions results in,
\bea
U_{1}(x,t) & = &
\frac{e^{-t}(8t)^{1/4}}{2\sqrt{2\pi}}\frac{e^{\sqrt{-8t\ln x}}
x}{(-\ln x)^{3/4}}
\left[1+\sqrt{\frac{2t}{-\ln x}}\left(1-\frac{3}{4t}\right)\right] \\
U_{2}(x,t) & = &
\frac{(1-e^{-t})}{(2t)^{1/4}\sqrt{\pi}}e^{\sqrt{-8t\ln x}} x(-\ln
x)^{1/4}
\left[1+\frac{1}{2}\sqrt{\frac{2t}{-\ln
x}}\right.\times\nonumber\\ & &
\times\left.\left(\frac{1}{2}+\frac{1}{16t}\right)\right]
\eea

\paragraph{Intermittency indices.}

Applying the expression for the intermittency indices to the
asymptotic expansions found above, we have,
\bea
\varphi_{2} & = & 1+\frac{1}{-\ln x} - \frac{1}{(-\ln
x)^{3/2}}\left(\frac{-11}{8}\sqrt{2t} + \frac{25}{64}
\sqrt{\frac{2}{t}}\right) + \cdots \\
\varphi_{n} & = & (n-1) + \frac{(n-1)}{2(-\ln x)} -
\frac{A}{\sqrt{t}(-\ln x)^{3/2}} + \cdots\\
A & = &
(n-1)(n-2)\frac{2^{\frac{n+1}{2}}}{16}t^{2-n/2}-\frac{(n-1)t}{\sqrt{2}}-\frac{\sqrt{2}}{16}(n^{2}-4n-3)
\nonumber
\eea
We see, then, that for $\sqrt{2t}\ll \sqrt{-\ln x}$ and $n\ll
-\ln x$  we have intermittency with a monofractal spectrum of
indices, with logarithmic corrections that, to first order, do not
depend on $t$.

\subsection{Singular Splitting Kernel.}

For the ``twin'' model \cite{chiuhwa},  we have $P(z)=\delta(z-1/2)$,
$\wt{P}=1/2^{s}$. We shall proceed here as before, by expanding the exponential
in (\ref{eq:10}),
\bea
\wt{d}_{1}(s,t) & = & e^{  -t}\sum_{n=0}^{\infty}\frac{(2t)^{  n}}{n!}
		\left(\frac{1}{2^{ s-1}}\right)^{  n}\\
\wt{d}_{2}(s,t) & = & 2(1-e^{  -t})\sum_{n=0}^{\infty}\frac{(2t)^{  n}}{n!}
		\left(\frac{1}{2^{ s-1}}\right)^{  n+1}\\
\wt{d}_{n}(s,t) & = & e^{  (n-2)t}(1-e^{  -t})^{  n-1}\sum_{m=1}^{n-1}
		C_{m}^{n-1}2^{  m}\sum_{k=0}^{\infty}\frac{(2t)^{  k}}{k!}\times\nonumber\\
	         &    &\times	\left(\frac{1}{2^{ s-1}}\right)^{  k+m}
\eea
and applying the inverse Mellin transform term by term to find,
\bea
d_{1}(x,t) & = & e^{  -t}\sum_{n=0}^{\infty}\frac{(2t)^{  n}}{n!}
		\delta\left(x-\frac{1}{2^{n}}\right)\\
d_{2}(x,t) & = & 2(1-e^{  -t})\sum_{n=0}^{\infty}\frac{(2t)^{  n}}{n!}
		\delta\left(x-\frac{1}{2^{n+1}}\right)\\
d_{n}(x,t) & = & e^{  (n-2)t}(1-e^{  -t})^{  n-1}\sum_{m=1}^{n-1}
		C_{m}^{n-1}2^{  m}\sum_{k=0}^{\infty}\frac{(2t)^{  k}}{k!}\times\nonumber\\
	  &     &	\times\delta\left(x-\frac{1}{2^{k+m}}\right)
\eea
For  the factorial moment densities we have,
\bea
U_{1}(x,t) & = & e^{  -t}\sum_{n=0}^{\infty}\frac{(2t)^{  n}}{n!}
	H\left(x-\frac{1}{2^{n}}\right)  =   e^{  -t}\sum_{n=[-\log_{2}
x]}^{\infty}\frac{(2t)^{  n}}{n!}\\
U_{2}(x,t) & = & 2(1-e^{  -t})\sum_{n=0}^{\infty}\frac{(2t)^{  n}}{n!}
	H\left(x-\frac{1}{2^{n+1}}\right) \nonumber\\
                 & =  & 2(1-e^{  -t})\sum_{n=[-\log_{2} x]-1}
	^{\infty}\frac{(2t)^{  n}}{n!}\\
U_{n}(x,t) & = & e^{  (n-2)t}(1-e^{  -t})^{  n-1}\sum_{m=1}^{n-1}
		C_{m}^{n-1}2^{  m}\sum_{k=0}^{\infty}\frac{(2t)^{  k}}{k!}
		H\left(x-\frac{1}{2^{k+m}}\right)\nonumber\\
                & = &  e^{  (n-2)t}(1-e^{  -t})^{  n-1}\sum_{m=1}^{n-1}
		C_{m}^{n-1}2^{  m}\sum_{k=[-\log_{2} x]-m}^{\infty}\frac{(2t)^{  k}}{k!}
\eea
where $[z]$ is the smallest integer $n$ such that $n\geq z$ and we denoted $H$
the right
continuous Heaviside step function.

We shall use now the fact that,
\[
\sum_{n=N}^{\infty}\frac{r^{n}}{n!} = \frac{e^{  r}}{\Gamma(N)}\gamma(N,r)
\]
$\gamma$ being an incomplete gamma function \cite{abra}, so that,
\bea
U_{1}(x,t) & = & e^{  t} \frac{\gamma([-\log_{2} x], 2t)}{\Gamma([-\log_{2}
x])}  \\
U_{2}(x,t) & = & 2e^{  2t}(1-e^{  -t}) \frac{\gamma([-\log_{2} x]-1,
2t)}{\Gamma([-\log_{2} x]-1)}  \\
U_{n}(x,t) & = & e^{  nt}(1-e^{  -t})^{  n-1}\sum_{m=1}^{n-1} C_{m}^{n-1} 2^{
m}
	 \frac{\gamma([-\log_{2} x]-m, 2t)}{\Gamma([-\log_{2} x]-m)}
\eea

\subsubsection{Large $t$.}

For fixed $x$ and $t\rightarrow\infty$ the ratio of  gamma functions goes to 1
and we have KNO scaling,
$U_{q}/U_{1}^{q}\rightarrow q!$, by definition of the coefficients
$C_{m}^{n-1}$, as in the previous case.

\subsubsection{Small $x$.}

The functions $U_{q}$ have a stairway shape superimposed to the general profile
obtained by eliminating
the integer part from the arguments. We shall work with this smoothed form in
what follows. The
density $U_{q}(U_{1})$ is unaltered by these device, as can be explicitly
checked, and
it is in this density that we are interested.

We shall now consider the form of these functions for large
$y\equiv-\log_{2}(x)$. To that end we use
the following asymptotic expansion  for large values of the parameter,
obtained by a saddle-point approximation,
\bea
\gamma(y,x) & = & \frac{x^{  y} e^{  -x}}{y-1}\left[
1+\frac{x-1}{y}+\frac{x(x-2)}{y^{  2}}+\cdots\right] \\
\Gamma(y) & = & \sqrt{2\pi} e^{  -y} y^{  y-1/2} \left[ 1+\frac{1}{12
y}+\frac{1}{288 y^{2}}+\cdots\right]
\eea
This way, we obtain, to first order in $1/y$,
\bea
\log_{2} U_{1} & = & \mbox{} - y\log_{2}
(y)+y\log_{2}(2te)-\frac{1}{2}\log_{2}(y)+\log_{2}\left(\frac{e^{  -t}}%
                                     {\sqrt{2\pi}}\right)+\nonumber\\
                        &    & \mbox{} +
\frac{2t-1/2}{y}+O\left(\frac{1}{y^2}\right) \\
\log_{2} U_{2} & = & \mbox{} - y\log_{2}
(y)+y\log_{2}(2te)+\frac{1}{2}\log_{2}(y)+\log_{2}\left(\frac{1-e^{  -t}}%
                                     {\sqrt{2\pi}t}\right)+\nonumber\\
                       &    & \mbox{} +
\frac{2t-1/2}{y}+O\left(\frac{1}{y^2}\right)
\eea
The case of $U_{n}$ is slightly more complicated since  it is a sum of several
terms. Assuming
$y\gg 2t+n$ we have, for $k\leq n$,
\be
\frac{\gamma(y-k,2t)}{\Gamma(y-k)}\frac{\Gamma(y-k+1)}{\gamma(y-k+1,2t)}\cong
\frac{y-k}{2t}\gg 1
\ee
and then we can retain only the leading and next-to-leading terms $n=n-1$,
$n-2$, so that,
\bea
U_{n} & \cong  & e^{  nt} (1-e^{  -t})^{  n-1} \left[ 2^{  n-1}
\frac{\gamma(y-n+1,2t)}{\Gamma(y-n+1)} +
\right.\nonumber\\
  &      &\mbox{} \left. + \frac{(n-1)(n-2)}{2}  2^{  n-2}
\frac{\gamma(y-n+2,2t)}{\Gamma(y-n+2)} + \cdots\right]
\eea
We shall use a first order approximation for  gamma functions in the first term
and  zeroth order for the
ones in  the second.  Up to higher order terms we finally have,
\bea
\log_{2} U_{n} & =  & -y\log_{2} y+y\log_{2} (2te) + \left(
n-\frac{3}{2}\right)\log_{2} y + cst.+\nonumber\\
                        &     & \mbox{} +\frac{1}{y}
   \left[ 2t -\frac{13}{12} + \frac{(n-1)(n-2)}{4} 2t -\frac{n(n-3)}{2}\right]
\eea
Proceeding as in the previous case, we arrive at the following expressions for
the intermittency indices,
\bea
\varphi _{2} & =  & 1 +
\frac{1}{y(\ln(y)-\ln(2t))}+O\left(\frac{1}{y^2}\right)\\
\varphi _{n} & =  & (n-1) +
\frac{n-1/2}{y(\ln(y)-\ln(2t))}+O\left(\frac{1}{y^2}\right)
\eea
Taking into account that $y = -\log_{2}(x)$, we see that for $-\log_{2}\gg
2t+n$ we have the same spectrum of asymptotic intermittency indices as
in the case of a uniform $P(x)$, with logarithmic corrections
independent of $t$ to first order.

\section{Final Remarks.}

In the previous sections we considered two particular cases of
one-species branching processes of the pure-birth type with binary
fission, characterized by a uniform and a Dirac-delta splitting
kernel, respectively. We exploited the fact that for these special
kernels explicit
solutions to the evolution equations for inclusive distributions can
be found. We have shown that for a fixed resolution in $x$, and
$t\rightarrow\infty$ a KNO scaling regime is reached, while for $t$
fixed and $x\rightarrow 0$ factorial moments grow without bound,
giving asymptotic intermittency indices $\varphi_{q}=q-1$ in both cases.
Leading corrections to this behavior for finite $x$ are proportional
to $1/\ln x$, and independent of $t$. Furthermore, they are positive,
{\em i.\frenchspacing e.\frenchspacing,} the asymptotic linear region
is approached from above.

The intermittent regime is defined as $\sqrt{2t}\ll \sqrt{-\ln x}$ for
the uniform and as $2t\ll -\log_{2} x$ for the singular kernel, for
the first few indices with $n\sim 1$. It
follows that $x\ll\langle x\rangle(t)$ is a necessary condition, which
seems difficult to fulfill in practical applications such as numerical
simulations. In \cite{chiuhwa} a Monte Carlo experiment with three
branching models is reported, of which only the ``twin'' model has
been considered here. Their results, as shown in Figure 2 of
\cite{chiuhwa}, are not inconsistent with our prediction for $n\leq
4$, within
numerical uncertainties in their computation, although it should be
noticed that for $q\geq 4$ their indices are systematically larger than
$q-1$, which is the maximum possible value \cite{bialzal}.

Even though our approach does not appear to be easy to extend to more
complicated functional forms of the kernel $P(x)$, we conjecture that
the spectrum of intermittency indices is independent of it within the
class of models considered in this paper. Our results may be useful to
tune up numerical studies of these models, along the lines of
\cite{chiuhwa}. We believe that a richer intermittent behavior, as
observed in experimental data,  may be
obtained  with infrared divergent kernels, about which results
have already been obtained \cite{brax2}.

\section*{Acknowledgments.}

I would like to thank Prof.\ R.\ Peccei for encouragement and support
during completion of this work. I benefitted also from many discussions
with Dr.\ P.\ Van Driel.

This work was supported by ICSC-World Laboratory.

\setcounter{equation}{0}
\renewcommand{\theequation}{A\arabic{equation}}

\section*{Appendix.}

We present here some basic equations and definitions from the theory
of branching processes as used in sections 2 and 3.

The starting point is the generating functional of the process,
$\phi[W,t;f]$ \cite{webber,cvitanovic}, which is a function of the
evolution parameter $t$ and of the energy $W$ of the initial particle,
and a functional of the dummy function $f(\omega )$. Probability
densities will be obtained by functional differentiation with respect
to $f$. Notice that $\omega$ has dimension of energy, like $W$.

The evolution of $\phi$ is given by the equations
\cite{webber,cvitanovic},
\bea
\frac{\partial\phi}{\partial t}[W,t;f] & = & \int_{0}^{1} dz\:
P(z)\left(\rule{0ex}{3ex}\phi[zW,t;f] \phi[(1-z)W,t;f] - \phi[W,t;f]\right) \\
\frac{\partial\phi}{\partial t}[W,t;f] & = & \int_{0}^{1} dz\:
\int_{0}^{1} d\xi\: P(z) W\frac{\delta\phi [W,t;f]}{\delta f(\xi
W)}\times  \nonumber \\
&  & \times\left(\rule{0ex}{3ex}f(z\xi W) f((1-z)\xi W)
 - f(\xi W))\right)
\eea
The first being the backward and the second the forward Kolmogorov
evolution equations \cite{feller}. These two equations have the same
set of solutions \cite{feller}, and only the forward equation is used
above. We use the initial condition $\phi[W,t=0;f] = f(W)$,
corresponding to one particle in the initial state.

Inclusive probability densities are defined as,
\be
D_{n}(x_{1},\ldots,x_{n};t) = W^{n}\left. \frac{\delta^{n}\phi
[W,t;f]}{\delta f(x_{n}W)\ldots\delta f(x_{1}W)}\right|_{f=1}
\ee
where we have already used the fact that since the splitting kernel
$P(x)$ depends only on energy fraction, and not on energy, the process
is scale invariant and probability densities depend only on
$x_{1},\ldots,x_{n}$ \cite{cvitanovic}. Furthermore, we see from this
definition that $D_{n}$ is completely symmetric in its arguments.

 From the evolution equations for $\phi$ we can derive the
corresponding equations for $D_{n}$. The forward one is given by
(\ref{eq:for}) and the backward equation is,
\bea
\lefteqn{\frac{\partial D_{n}}{\partial t}
(x_{1},\ldots,x_{n};t)=}\nonumber\\ & & \int_{0}^{1} dz\:
P(z)\left(\sum_{\pi(n,k)} \frac{1}{z^{k}} D_{k}\left(
\frac{x_{i_{1}}}{z},\ldots,\frac{x_{i_{k}}}{z};t\right)\right.\times
\nonumber\\
& & \times\left.\frac{1}{(1-z)^{n-k}} D_{n-k} \left(
\frac{x_{i_{k+1}}}{z},\ldots,\frac{x_{i_{n}}}{z};t\right) -
D_{n}(x_{1},\ldots,x_{n};t) \rule{0ex}{5ex}\right)
\eea
where the sum runs over all possible partitions $\pi(n,k)$ of
$(1,\ldots,n)$ into two sets $(i_{1},\ldots,i_{k})$,
$(i_{k+1},\ldots,i_{n})$ such that $i_{1}>i_{2}>\cdots >i_{k}$ and
$i_{k+1}>\cdots >i_{n}$, with $k=0,\ldots,n$; and we used the
relation,
\be
W\left. \frac{\delta\phi [zW,t;f]}{\delta f(xW)}\right|_{f=1} =
\frac{1}{z} D_{1}\left(\frac{x}{z};t\right)
\ee
and its generalizations to higher order densities $D_{n}$.

Analogously, exclusive distributions are defined by,
\be
E_{n}(x_{1},\ldots,x_{n};t) = W^{n}\left. \frac{\delta^{n}\phi
[W,t;f]}{\delta f(x_{n}W)\ldots\delta f(x_{1}W)}\right|_{f=0}
\ee
and satisfy,
\bea
\lefteqn{\frac{\partial E_{n}}{\partial t}
(x_{1},\ldots,x_{n};t)= \int_{0}^{1} dz\:
P(z)\left(\sum_{\pi(n,k)} \frac{1}{
{n\choose k}z^{k}(1-z)^{n-k}}\right.\times}\nonumber\\
& & \left.
E_{k}\left(\frac{x_{i_{1}}}{z},\ldots,\frac{x_{i_{k}}}{z};t\right)
E_{n-k} \left(\frac{x_{i_{k+1}}}{z},\ldots,\frac{x_{i_{n}}}{z};t\right) -
E_{n}(x_{1},\ldots,x_{n};t) \rule{0ex}{5ex}\right)\\
\lefteqn{\frac{\partial E_{n}}{\partial t}(x_{1},\ldots,x_{n},t) =
- n E_{n}(x_{1},\ldots,x_{n},t) + \frac{2}{n}\sum_{k>j=1}^{n}
P\left(\frac{x_{j}}{x_{j}+x_{k}}\right)\times}
\nonumber \\
& &
\frac{1}{x_{j}+x_{k}} E_{n-1}(x_{1},\ldots,\underbrace{x_{j}+x_{k}}_{j},
\ldots,\widehat{x_{k}},\ldots,x_{n},t)
\eea
The initial condition is given by
$E_{n}(x_{1},\ldots,x_{n};t=0)=\delta_{1n}\delta(x_{1}-1)$.

The relation between $D_{n}$ and $E_{n}$ may be explicitly written as,
\bea
D_{n}(x_{1},\ldots,x_{n};t) & = & \sum_{N=n}^{\infty} N(N-1)\cdots
(N-n+1) \times\nonumber\\
&  & \int_{0}^{1} dy_{N=1}\cdots dy_{N}\:
E_{n}(x_{1},\ldots,x_{n},y_{1},\ldots,y_{n};t)
\eea
where, once again, we have made use of the symmetry with respect to
permutations of the arguments of these distributions. Notice that our
definition of $E_{n}$ is slightly different from that of
\cite{cvitanovic}, since we normalize $E_{n}$ to the multiplicity
distribution,
\be
\int_{0}^{1} dx_{1}\cdots dx_{n}\: E_{n}(x_{1},\ldots,x_{n};t) =
P_{n}(t)
\ee
from whence the normalization of $D_{n}$ to factorial moments
results.

It is not difficult to convince oneself that solutions to the
evolution equations have the form,
\be
E_{n}(x_{1},\ldots,x_{n};t) = P_{n}(t) K_{n}(x_{1},\ldots,x_{n})
\ee
where $K_{n}$ is defined by this equation. The evolution equations for
$P_{n}(t)$, the multiplicity distribution, can be obtained by
integration of the equations for $E_{n}$. Since these are well-known
\cite{giova,bouzas}, we shall quote only the solutions,
\bea
P_{n}(t) & = & \frac{1}{\langle n\rangle}\left(\frac{\langle n\rangle
-1}{\langle n\rangle}\right)^{n-1} \\
\langle n\rangle & = & e^{t}
\eea
 From this expression, and the evolution equations for $E_{n}$,
recurrence relations (backward and forward) for
$K_{n}(x_{1},\ldots,x_{n})$ can be found. We shall only write the
forward one,
\bea
K_{n}(x_{1},\ldots,x_{n}) & = & \frac{2}{n(n-1)}\sum_{k>j=1}^{n}
P\left(\frac{x_{j}}{x_{j}+x_{k}}\right)
\frac{1}{x_{j}+x_{k}} \times\nonumber \\
&  & K_{n-1}(x_{1},\ldots,\underbrace{x_{j}+x_{k}}_{j},
\ldots,\widehat{x_{k}},\ldots,x_{n})
\eea
with $K_{1}(x)=\delta(x-1)$. From this relation and $P(x)=P(1-x)$ it
follows that,
\be
\int_{0}^{1} dx_{1}\cdots dx_{n}\: x_{1} K_{n}(x_{1},\ldots,x_{n}) =
\frac{1}{n}
\ee
and then,
\be
\langle x\rangle \equiv \sum_{n=1}^{\infty}\int_{0}^{1} dx_{1}\cdots
dx_{n}\: x_{1} E_{n}(x_{1},\ldots,x_{n};t) = \frac{\ln \langle
n\rangle}{\langle n\rangle - 1}
\ee
independently of the detailed form of $P(x)$.

In \cite{hwa}, it is shown that a branching process satisfying,
\be
E_{n}(x_{1},\ldots,x_{n};t) = \prod_{i=1}^{n}
f(x_{i};t)\delta\left(1-{\sum_{j=1}^{n}} x_{j}\right)
\ee
is not intermittent. From the factorization property discussed above,
we see that such a branching process does not exist, at least within
the class of models considered here.


\begin{thebibliography}{00}
\bibitem{bialpes} A.\ Bia{\l}as and R.\ Peschanski, Nucl.\ Phys.\ {\bf B273},
(1986),703; {\bf B308}, (1988), 857.
\bibitem{bialrev} A.\ Bia{\l}as, Nucl.\ Phys.\ {\bf A545}, (1992), 285c.
\bibitem{pesrev} R. Peschanski, Int.\ Jou.\ Mod.\ Phys.\ {\bf A6},
                (1991), 3681.
\bibitem{schmitz} N.\ Schmitz, Proc.\ XXI Int.\ Symp.\ Multiparticle
Dynamics, Wuhan, China, 1991 (W.\ Yuantang and L.\
Lianshou Eds.\frenchspacing, World Scientific, Singapore,
1992)p.\frenchspacing 377.
\bibitem{buschbeck} B.\ Buschbeck, Proc.\  XXVI Rencontre de
Moriond, Les Arcs, France, 1991 (Tran Thanh Van Ed.\frenchspacing,
Frontiers,  J. Gif sur Yvette, 1991)p.\frenchspacing 299.
\bibitem{brax} P.\ Brax and R.\ Peschanski, Int.\ Jou.\ Mod.\ Phys.\
{\bf A10}, (1992), 709.
\bibitem{zal} A.\ Bia{\l}as and K.\ Zalewski, Phys.\ Lett.\ {\bf B238},
(1990), 413.
\bibitem{bhar} A.\ Bharucha-Reid, Elements of the theory of Markov proces%
               ses and their applications (Mc.\ Graw-Hill, New
York, 1960).
\bibitem{polya} A.\ Polyakov, Sov.\ Phys.\ JETP {\bf 32}, (1971), 296;
                {\bf 33}, (1971), 850.
\bibitem{orfa} S.\ Orfanidis and V.\ Rittenberg, Phys.\ Rev.\ {\bf D10},
               (1974), 2892.
\bibitem{giova} A.\ Giovannini, Nucl.\ Phys.\ {\bf B161}, (1979), 429.
\bibitem{hwa} R.\ Hwa, Nucl.\ Phys.\ {\bf B238}, (1989), 59.
\bibitem{brax1} P.\ Brax and R.\ Peschanski, SACLAY Preprint
SPHT-92-005, 1992.
 \bibitem{brax2} P.\ Brax, J.\ Meunier and R.\ Peschanski,
INLN Preprint 93-01, 1993.
\bibitem{pesc}Y.\  Gabellini, J.\ Meunier and R.\ Peschanski,
Z.\ Phys.\ {\bf C55}, (1992), 455.
\bibitem{koba} Z.\ Koba, H.\ Nielsen and P.\ Olesen, Nucl.\ Phys.\
               {\bf B40}, (1972), 317.
\bibitem{bialzal} A.\ Bia{\l}as and K.\ Zalewski, Phys.\ Lett.\   {\bf
B228},(1989),155.
\bibitem{alta} G.\ Altarelli and G.\ Parisi, Nucl.\ Phys.\ {\bf B126},
(1977), 298.
\bibitem{gribo} V.\ Gribov and L.\ Lipatov, Sov.\ J.\ Nucl.\ Phys.\
                {\bf 15}, (1972), 675.
\bibitem{webber} B.\ Webber, Ann.\ Rev.\ Part.\ Phys.\ {\bf 36},
(1986), 253.
\bibitem{cvitanovic} P.\ Cvitanovi\'c, P.\ Hoyer and K.\ Zalewski,
                     Nucl.\ Phys.\ {\bf B176}, (1980), 429.
\bibitem{sjostrand} M.\ Bengtson and T.\ Sj\"ostrand, Nucl.\ Phys. {\bf B289},
                    (1987), 810.
\bibitem{bouzas} A.\ Bouzas, Z.\ Phys.\ {\bf C}, (1993), to appear.
\bibitem{pespes} R.\ Peschanski, Proc.\ Santa Fe
Workshop on Intermittency in High energy Collisions, Los Alamos, USA,
1990 (F.\ Cooper, R.\
Hwa and I.\ Sarcevic Eds.\frenchspacing, World Scientific, Singapore,
1991)p.\frenchspacing 158.
\bibitem{ochs} W.\ Ochs, Z.\ Phys.\ {\bf C50}, (1991), 339.
\bibitem{abra} M.\ Abramowitz and I.\ Stegun Eds.\frenchspacing,
Handbook of Mathematical Functions (National Bureau of Standards,
Applied Mathematics Series v.\frenchspacing 55, 1968).
\bibitem{erdelyi} A.\ Erd\'elyi Ed.\frenchspacing, Higher
Trascendental Functions (R.\ Krieger, Malabar,
Florida, 1981).
\bibitem{chiuhwa} C.\ Chiu and R.\ Hwa, Proc.\ Santa Fe
Workshop on Intermittency in High energy Collisions, Los Alamos, USA,
1990 (F.\ Cooper, R.\
Hwa and I.\ Sarcevic Eds.\frenchspacing, World Scientific, Singapore,
1991)p.\frenchspacing 202.
\bibitem{feller} W.\ Feller, An Introduction to Probability
Theory and its Applications, Vol.\  2 (John Wiley, New York,
1971).
\end{thebibliography}
\end{document}